\documentclass[utf8]{frontiersinFPHY_FAMS} 
\setcitestyle{square} 
\usepackage{url,hyperref,microtype,subcaption}
\usepackage[onehalfspacing]{setspace}
\usepackage{amsmath}
\usepackage{amssymb}

\def\keyFont{\fontsize{8}{11}\helveticabold }
\def\firstAuthorLast{Siv\'y {et~al.}} 
\def\Authors{D\'avid Siv\'y\,$^{1,\dagger}$, Katar\'ina Karl'ov\'a\,$^{2}$ 
and Jozef Stre\v{c}ka\,$^{1,\dagger,*}$}
\def\Address{$^{1}$Department of Theoretical Physics and Astrophysics, Faculty of Science, 
Pavol~Jozef~\v{S}af\'{a}rik University, Ko\v{s}ice, Slovakia \\
$^{2}$Laboratoire de Physique Th\'eorique et Mod\'elisation, CY Cergy Paris Universit\'e, Cergy-Pontoise, France}
\def\corrAuthor{Corresponding Author}
\def\corrEmail{jozef.strecka@upjs.sk}

\begin{document}
\onecolumn
\firstpage{1}

\title[Quantum Magnetism in Fe$_2$Cu$_2$ Polymeric Branched Chains]{Quantum Magnetism in Fe$_2$Cu$_2$ Polymeric Branched Chains: Insights from Exactly Solved Ising-Heisenberg Model} 

\author[\firstAuthorLast ]{\Authors} 
\address{} 
\correspondance{} 

\extraAuth{}

\maketitle

\begin{abstract}
The spin-1/2 Ising-Heisenberg branched chain, inspired by the magnetic structure of three isostructural polymeric coordination compounds [(Tp)$_{2}$Fe$_{2}$(CN)$_{6}$X(bdmap)Cu$_{2}$(H$_{2}$O)] $\cdot$ H$_2$O to be further denoted as Fe$_2$Cu$_2$ (Tp = tris(pyrazolyl)hydroborate, bdmapH = 1,3-bis(dimethylamino)-2-propanol, HX = acetic acid, propionic acid or trifluoroacetic acid), is rigorously studied using the transfer-matrix method. The overall ground-state phase diagram reveals three distinct phases: a quantum antiferromagnetic phase, a quantum ferrimagnetic phase and a classical ferromagnetic phase. In the zero-temperature magnetization curve, two quantum ground states are manifested as intermediate plateaus at zero and half of the saturation magnetization, while the magnetization reaches its saturated value within the classical ferromagnetic phase. The bipartite entanglement between nearest-neighbor Heisenberg spins is more pronounced in the quantum ferrimagnetic phase compared to the quantum antiferromagnetic phase due to a fully polarized nature of the Ising spins. A reasonable agreement between theoretical predictions for the spin-1/2 Ising-Heisenberg branched chain and experimental data measured for a temperature dependence of the magnetic susceptibility and a low-temperature magnetization curve suggests strong antiferromagnetic coupling between nearest-neighbor Cu$^{2+}$-Cu$^{2+}$ magnetic ions and moderately strong ferromagnetic coupling between nearest-neighbor Cu$^{2+}$-Fe$^{3+}$ magnetic ions in the polymeric compounds Fe$_2$Cu$_2$. A thermal entanglement between nearest-neighbor Cu$^{2+}$-Cu$^{2+}$ magnetic ions persists up to a relatively high threshold temperature $T \approx 224$~K and undergoes a transient magnetic-field-driven strengthening. 
\tiny
 \keyFont{ \section{Keywords:} Ising-Heisenberg Branched Chain, Quantum and Thermal Entanglement, Fractional Magnetization Plateau, Magnetocaloric Effect, Heterobimetallic Coordination Polymers, Exact Results} 
\end{abstract}

\section{Introduction}

Although one-dimensional (1D) quantum spin models may seem  relatively simple at first glance, achieving an exact solution for these lattice-statistical models is often accompanied with formidable mathematical difficulties stemming from noncommuting algebra \cite{mat93}. As a matter of fact, there exists only a few paradigmatic examples of exactly solved 1D quantum spin models such as the quantum spin-1/2 Heisenberg chain solved by the Bethe-ansatz method \cite{hei28,bet31,hul38,gri64} or the quantum spin-1/2 XY model solved using the Jordan-Wigner fermionization approach \cite{lie61,kat62}. It should be stressed that the exploration of 1D quantum spin models is of particular importance, because several low-dimensional magnetic materials may have negligible exchange interactions in the remaining two spatial dimensions \cite{mil02}. It has been demonstrated that several 1D spin chains may display intriguing quantum features such quantized magnetization plateaus \cite{ayd04} and quantum entanglement \cite{han17}.

1D Ising-Heisenberg models featuring regularly alternating Ising and Heisenberg spins are conversely relatively easily exactly solvable using the transfer-matrix method. In spite of their simplicity, these models still offer a plausible framework for theoretical modeling of a certain class of insulating quantum magnetic materials. The first paradigmatic experimental realization of the spin-1/2 Ising-Heisenberg chain composed from regularly alternating Ising spins and Heisenberg spin trimers was found by van den Heuvel and Chibotaru during their analysis of magnetic properties of the trimetallic 3d-4d-4f coordination polymer [\{(CuL)$_2$Dy\}\{Mo(CN)$_8$\}]$\cdot$CH$_3$CN$\cdot$H$_2$O, where L$^{2-}$ = N,N’-propylenebis(3-methoxysalicylideneiminato) \cite{heu10,bel14}. A simpler experimental realization of the spin-1/2 Ising-Heisenberg chain, composed from regularly alternating Ising and Heisenberg spins, was discovered in the bimetallic 3d-4f coordination polymer Dy(NO$_3$)(DMSO)$_2$Cu(opba)(DMSO)$_2$ (DMSO = dimethylsulfoxide, opba = orthophenylenebisoxamato) \cite{str12,han13,tor18}. The low-temperature  magnetization data of another 3d-4f heterobimetallic coordination polymer [{Dy(hfac)$_2$(CH$_3$OH)}$_2${Cu(dmg)(Hdmg)}$_2$]$_n$ (H$_2$dmg = dimethylglyoxime; Hhfac = 1,1,1,5,5,5-hexafluoropentane-2,4-dione) have been successfully interpreted within the spin-1/2 Ising-Heisenberg orthogonal-dimer chain \cite{str20,gal22}. It is worthwhile to remark that several other exactly solved Ising-Heisenberg spin chains have provided unbiased descriptions of various intriguing quantum phenomena including quantized magnetization plateaus \cite{can06,str14}, enhanced magnetocaloric effect \cite{str14,qiy14}, partition function zeros \cite{hov16}, quantum teleportation \cite{roj17}, unconventional ‘fire-and-ice’ ground states \cite{rot18}, pseudo-transitions \cite{car19}, quantum coherence \cite{fre19}, quantum nonlocality \cite{bhu21}, quantum Fisher information \cite{ren22} as well as nanostructures \cite{ma22,ma23}.

In recent years, considerable attention has been directed towards investigating magnetic and quantum features of Ising-Heisenberg branched chains. Notably, the mixed spin-(1/2,5/2,1/2) Ising-Heisenberg branched chain has demonstrated its utility in theoretical modeling the magnetic characteristics of the heterotrimetallic coordination polymer [CuMn(L)][Fe(bpb)(CN)$_2$]ClO$_4$ $\cdot$ H$_2$O \cite{sou19,sou20}. The mixed spin-(1/2,5/2,1/2) Ising-Heisenberg branched chain provided a satisfactory explanation of the magnetization and susceptibility behavior of this heterotrimetallic Fe-Mn-Cu coordination polymer. Furthermore, the Fe-Mn-Cu polymeric branched chain has been proposed as a promising platform for realizing quantum teleportation \cite{zhe19,ari21}, while also exhibiting robust quantum coherence \cite{ari21} and thermal entanglement \cite{zhe22}. On the other hand, the spin-1/2 Ising-Heisenberg branched chain inspired by the magnetic structure of the bimetallic coordination polymer [(Tp)$_2$Fe$_2$(CN)$_6$(OCH$_3$)(bap)Cu$_2$(CH$_3$OH) $\cdot$ 2CH$_3$OH $\cdot$ H$_2$O] (Tp = tris(pyrazolyl)hydroborate, bapH = 1,3-bis(amino)-2-propanol) theoretically predicts an intriguing breakdown of the intermediate one-half magnetization plateau in a low-temperature magnetization curve depending basically on the interaction anisotropy \cite{kar19,str22}.

\begin{figure*}[t]
\centering
\includegraphics[scale=0.5]{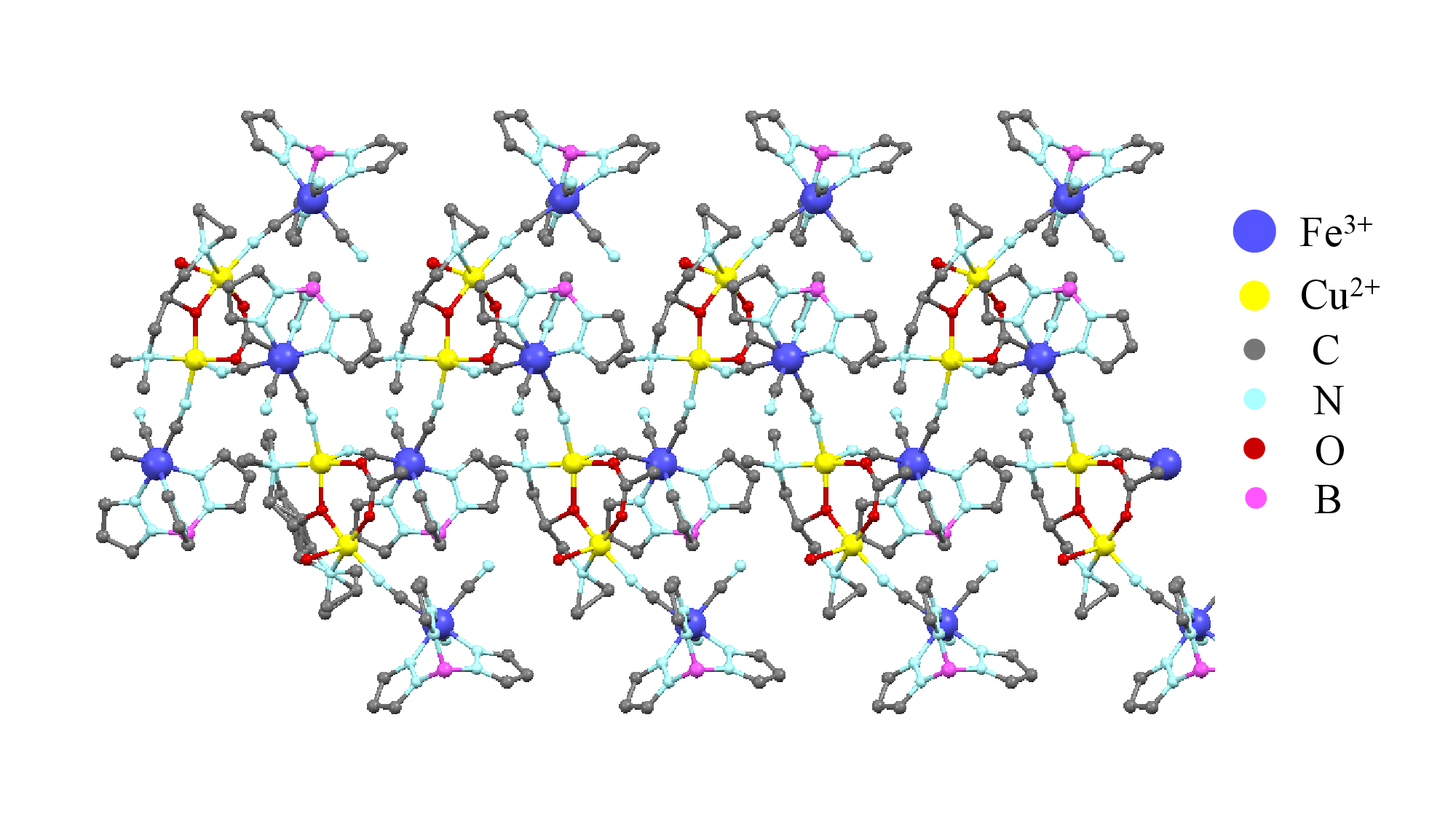}
\vspace{-1.3cm}
\caption{A part of the crystal structure of the polymeric coordination compound [(Tp)$_{2}$Fe$_{2}$(CN)$_{6}$(CH$_3$COO)(bdmap)Cu$_{2}$(H$_{2}$O)]  depicted according to crystallographic data reported in Ref. \cite{kan10}. Tp stands for tris(pyrazolyl) hydroborate and bdmapH is 1,3-bis(dimetylamino)-2-propanol.}
\label{fig1}
\end{figure*} 

\begin{figure}[t]
\centering
\includegraphics[scale=0.5]{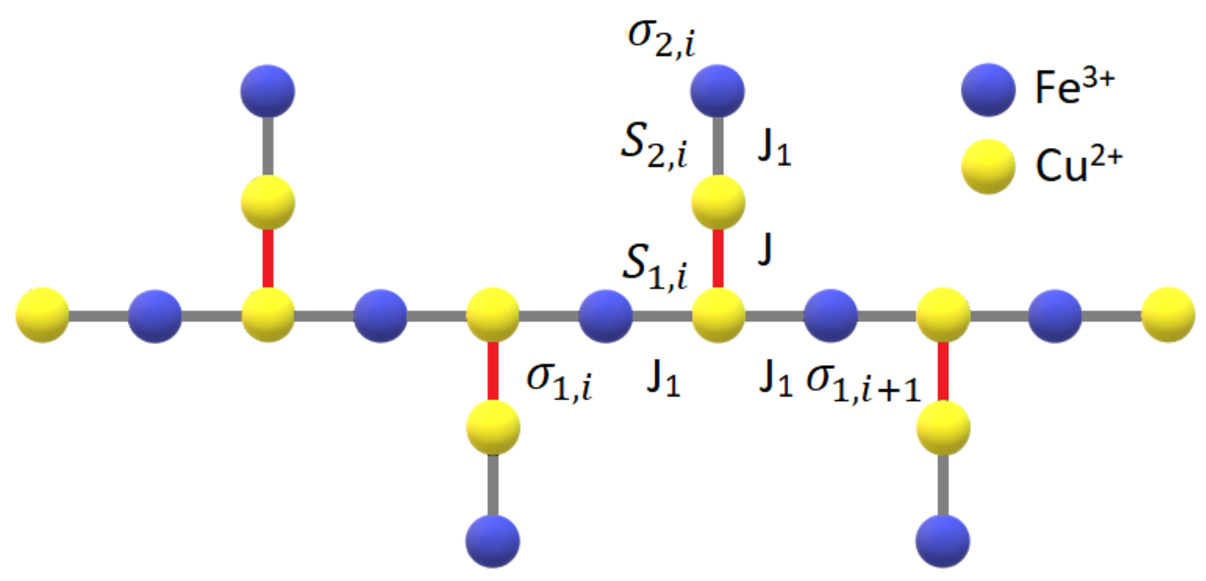}
\vspace{-0.3cm}
\caption{A schematic illustration of the spin-1/2 Ising-Heisenberg branched chain with the isotropic exchange interaction $J$ between nearest-neighbor Heisenberg spins and the anisotropic exchange interaction $J_{1}$ between the nearest-neighbor Heisenberg and Ising spins, respectively. The Heisenberg spins capture lattice positions of Cu$^{2+}$ magnetic ions (yellow circles), while the Ising spins capture lattice positions of Fe$^{3+}$ magnetic ions (blue circles).}
\label{fig2}
\end{figure}

In the present study, we will introduce a similar but structurally distinct spin-1/2 Ising-Heisenberg branched chain  to model the magnetic properties of three isostructural polymeric coordination compounds [(Tp)$_{2}$Fe$_{2}$(CN)$_{6}$X(bdmap)Cu$_{2}$(H$_{2}$O)] as reported in Ref. \cite{kan10} [Tp=tris(pyrazolyl) hydroborate, bdmapH=1,3-bis(dimetylamino)-2-propanol and HX is acetic acid, propionic acid or trifluoracetic acid]. The magnetic structure of these polymeric coordination compounds, further abbreviated as Fe$_2$Cu$_2$, is shown in Fig. \ref{fig1} and it can be identified as another variant of spin-1/2 Ising-Heisenberg branched chain. This characterization arises from the presence of magnetically isotropic spin-1/2 Cu$^{2+}$ magnetic ions ($S=1/2$) described by the notion of quantum Heisenberg spins, which is in contrast to highly magnetically anisotropic spin-1/2 Fe$^{3+}$ magnetic ions ($\sigma=1/2$) described by the notion of Ising spins. Consequently, the exchange coupling between nearest-neighbor Cu$^{2+}$-Cu$^{2+}$ magnetic ions will be assumed to be of the Heisenberg type, whereas the exchange coupling between nearest-neighbor Fe$^{3+}$-Cu$^{2+}$ magnetic ions will be treated as the Ising-type exchange interaction (see Fig. \ref{fig2}). 

The structure of this paper is as follows. In Section 2, we introduce the spin-1/2 Ising-Heisenberg branched chain along with a brief outline of the method employed for its calculation. Section 3 is dedicated to discussing the theoretical findings regarding the ground-state phase diagram, magnetization curves, entropy, and concurrence. In Section 4, we compare the exact theoretical results for the spin-1/2 Ising-Heisenberg branched chain with available experimental data for the magnetization and susceptibility of the polymeric compound Fe$_2$Cu$_2$ together with the respective theoretical prediction for the concurrence. Finally, Section 5 provides a summary of the most important scientific findings obtained in this study.

\section{Model and method}
\label{method}
The spin-1/2 Ising-Heisenberg branched chain schematically illustrated in Fig. \ref{fig2} can be defined through the following Hamiltonian:
\begin{eqnarray}
\hat{\mathcal H} \!=\! \sum_{i=1}^{N} \! \Bigl \{ J\!\left[\Delta \!\left(\!\hat{S}_{1,i}^{x} \hat{S}_{2,i}^{x} \!+\! \hat{S}_{1,i}^{y} \hat{S}_{2,i}^{y}\!\right) \!+\! \hat{S}_{1,i}^{z} \hat{S}_{2,i}^{z}\right] \!+\! 
J_{1} \!\left[\hat{S}_{2,i}^{z}\hat{\sigma}_{2,i}^{z} \!+\! 
\hat{S}_{1,i}^{z}\!\left(\hat{\sigma}_{1,i}^{z}\!+\!\hat{\sigma}_{1,i+1}^{z}\!\right)\!\right]\! 
\!-\! h_H \!\sum_{j=1}^{2} \! \hat{S}_{j,i}^{z} \!-\! h_I \!\sum_{j=1}^{2} \! \hat{\sigma}_{j,i}^{z} \! \Bigr\},
\!\!\!
\label{IsingHeiHam}
\end{eqnarray}
where $\hat{\sigma}_{j,i}^{z}$ and $\hat{S}^{\alpha}_{j,i}$ $(\alpha=x,y,z)$ label the Ising and Heisenberg spins ascribed to $\rm{Fe}^{3+}$ and $\rm{Cu}^{2+}$ magnetic ions, respectively. The coupling constant $J>0$ denotes the antiferromagnetic exchange interaction between the nearest-neighbor Heisenberg spin pairs capturing lattice positions of the dimeric $\rm{Cu}^{2+}-\rm{Cu}^{2+}$ units in the polymeric compounds Fe$_2$Cu$_2$, while the coupling constant $J_{1}>0$ ($J_{1}<0$) stands for the antiferromagnetic (ferromagnetic) exchange interaction between the nearest-neighbor Ising and Heisenberg spins capturing lattice positions of $\rm{Fe}^{3+}$ and $\rm{Cu}^{2+}$ magnetic ions, respectively. Furthermore, the Zeeman’s terms $h_H = g_{\rm Cu} \mu_{\rm B} B$ and $h_I = g_{\rm Fe} \mu_{\rm B} B$ account for the magnetostatic energy of magnetic moments ascribed to the Heisenberg and Ising spins in an external magnetic field $B$, $g_{\rm Cu}$ and $g_{\rm Fe}$ are the respective Land\'e g-factors and $\mu_{\rm B}$ is Bohr magneton. The parameter $\Delta$ measures a degree of the exchange anisotropy in the XXZ Heisenberg interaction and $N$ denotes the total number of unit cells. For simplicity, the periodic boundary conditions $\sigma_{1,N+1}=\sigma_{1,1}$ are assumed.

It is convenient to rewrite the total Hamiltonian \eqref{IsingHeiHam} as a sum of the cell Hamiltonians $\hat{\mathcal{H}}= \sum_{i=1}^N \hat{\mathcal{H}}_i$ with the following definition of the cell Hamiltonian $\hat{\mathcal{H}}_i$:
\begin{eqnarray}
\hat{{\mathcal H}_{i}} \!\!\!&=&\!\!\! J\left[\Delta(\hat{S}_{1,i}^{x} \hat{S}_{2,i}^{x} + \hat{S}_{1,i}^{y} \hat{S}_{2,i}^{y}) + \hat{S}_{1,i}^{z} \hat{S}_{2,i}^{z}\right] 
+ J_{1} \left[\hat{S}_{2,i}^{z} \hat{\sigma}_{2,i}^{z} + \hat{S}_{1,i}^{z}(\hat{\sigma}_{1,i}^{z} + \hat{\sigma}_{1,i+1}^{z})\right] \nonumber \\
\!\!\!&-&\!\!\! h_{H}(\hat{S}_{1,i}^{z} + \hat{S}_{2,i}^{z}) - \frac{h_{I}}{2}(\hat{\sigma}_{1,i}^{z} + \hat{\sigma}_{1,i+1}^{z} + 2\hat{\sigma}_{2,i}^{z}). 
\label{CellIsingHeiHam}
\end{eqnarray}
Owing to the fact that the cell Hamiltonians \eqref{CellIsingHeiHam} commute with each other $[\hat{\mathcal{H}}_i,\hat{\mathcal{H}}_j]=0$, the partition function of the spin-1/2 Ising-Heisenberg branched chain can be factorized into the following product:
\begin{eqnarray}
Z = \sum_{\{ \sigma_{1,i} \} } \prod_{i=1}^{N} \sum_{\sigma_{2,i}} 
{\rm Tr}_{i}~ \exp \left(- \beta \hat{{\mathcal H}_{i}}\right),
\label{IsiHeiPart}
\end{eqnarray} 
where $\beta=1/(k_{\rm{B}}T)$, $k_{\rm{B}}$ is the Boltzmann's constant, $T$ is the absolute temperature, ${\rm Tr}_{i}$ stands for a trace over degrees of freedom of the Heisenberg dimer $S_{1,i}-S_{2,i}$ from the $i$-th unit cell, $\sum_{\sigma_{2,i}}$ denotes a summation over two states of the Ising spin from the branching of the $i$-th unit cell, and the summation $\sum_{\{\sigma_{1,i}\}}$ runs over all possible configurations of the Ising spins from the main backbone of the branched chain. The partition function \eqref{IsiHeiPart} can be rewritten into a more suitable form for further calculations:
\begin{eqnarray}
Z = \sum_{\{ \sigma_{1,i} \} } \prod_{i=1}^{N} \exp\left[\frac{\beta h_{I}}{2}(\sigma_{1,i} + \sigma_{1,i+1})\right] \sum_{\sigma_{2,i}} \exp \left(\beta h_{I} \sigma_{2,i}\right) 
{\rm Tr}_{i}~ \exp \left(- \beta \hat{{\mathcal H}_{i}^{'}}\right),
\label{IsiHeiPartRE}
\end{eqnarray}   
where the reduced cell Hamiltonian $\hat{{\mathcal H}_{i}^{'}}$ is defined as:
\begin{eqnarray}
\hat{{\mathcal H}_{i}^{'}} = J[\Delta(\hat{S}_{1,i}^{x} \hat{S}_{2,i}^{x} + \hat{S}_{1,i}^{y} \hat{S}_{2,i}^{y}) + \hat{S}_{1,i}^{z} \hat{S}_{2,i}^{z}] + \sum_{j=1}^{2} h_{j,i} \hat{S}_{j,i}^{z}.
\label{HeiPartHam}
\end{eqnarray}
The reduced cell Hamiltonian $\hat{{\mathcal H}_{i}^{'}}$ given by Eq. (\ref{HeiPartHam}) can be alternatively viewed as the Hamiltonian of a spin-1/2 Heisenberg dimer in nonuniform effective fields $h_{j,i}$ involving the interaction with the adjacent Ising spins:
\begin{eqnarray}
h_{1,i} = J_{1}(\sigma_{1,i} + \sigma_{1,i+1}) - h_{H}, \qquad h_{2,i} = J_{1} \sigma_{2,i} - h_{H}.
\label{HeiEffFiel}
\end{eqnarray}
To perform a trace Tr$_i$ over degrees of freedom of the spin-1/2 Heisenberg dimer from the $i$-th unit cell in the relevant expression for the partition function \eqref{IsiHeiPartRE}, it is convenient to proceed to a matrix representation of the reduced cell Hamiltonian \eqref{HeiPartHam} in the standard basis:
\begin{eqnarray}
\left|\varphi_{1,i}\right\rangle = \left|\uparrow\right\rangle_{1,i} \left|\uparrow\right\rangle_{2,i}, 
\quad
\left|\varphi_{2,i}\right\rangle = \left|\uparrow\right\rangle_{1,i} \left|\downarrow\right\rangle_{2,i},
\quad
\left|\varphi_{3,i}\right\rangle = \left|\downarrow\right\rangle_{1,i} \left|\uparrow\right\rangle_{2,i}, 
\quad
\left|\varphi_{4,i}\right\rangle = \left|\downarrow\right\rangle_{1,i} \left|\downarrow\right\rangle_{2,i}.
\label{IsiHeiBasis}
\end{eqnarray}
The state vectors $\left|\uparrow\right\rangle_{j,i}$ and $\left|\downarrow\right\rangle_{j,i}$ are eigenvectors of the $z$-component of the Heisenberg spin $S_{j,i}^{z} = 1/2$ and $-1/2$, respectively. After solving the relevant eigenvalue problem $\left|\left\langle \varphi_{k,i}\right|\hat{{\mathcal H}_{i}^{'}}\left|\varphi_{l,i}\right\rangle-E_{j,i}{\rm \textbf{I}}\right|=0$ one gets four eigenvalues of the reduced cell Hamiltonian (\ref{HeiPartHam}): 
\begin{eqnarray}
E_{1,i} \!\!\!&=&\!\!\! \frac{J}{4} - \frac{h_{1,i}}{2} - \frac{h_{2,i}}{2}, \qquad \qquad \qquad \qquad \quad
E_{2,i} = \frac{J}{4} + \frac{h_{1,i}}{2} + \frac{h_{2,i}}{2}, \nonumber \\
E_{3,i} \!\!\!&=&\!\!\! - \frac{J}{4} + \frac{1}{2}\sqrt{\left(h_{1,i} - h_{2,i}\right)^{2} + \left(J \Delta\right)^{2}}, \qquad
E_{4,i} = - \frac{J}{4} - \frac{1}{2}\sqrt{\left(h_{1,i} - h_{2,i}\right)^{2} + \left(J \Delta\right)^{2}},
\label{IsiHeiEigenVal}
\end{eqnarray}
with the corresponding eigenvectors:
\begin{eqnarray}
\left|\psi_{1,i}\right\rangle\!\!\!&=&\!\!\!\left|\uparrow\right\rangle_{1,i} \left|\uparrow\right\rangle_{2,i},
\quad
\left|\psi_{2,i}\right\rangle= \left|\downarrow\right\rangle_{1,i} \left|\downarrow\right\rangle_{2,i},
\quad
\left|\psi_{3,i}\right\rangle=A_i^{+} \left|\uparrow\right\rangle_{1,i} \left|\downarrow\right\rangle_{2,i}
+A_i^{-}\left|\downarrow\right\rangle_{1,i} \left|\uparrow\right\rangle_{2,i},
\nonumber \\
\left|\psi_{4,i}\right\rangle\!\!\!&=&\!\!\!
A_i^{-}\left|\uparrow\right\rangle_{1,i} \left|\downarrow\right\rangle_{2,i}
-A_i^{+}\left|\downarrow\right\rangle_{1,i} \left|\uparrow\right\rangle_{2,i},  
\label{IsiHeiEigenVec}
\end{eqnarray}
where the probability amplitudes $A_i^{\mp}$ are given by:
\begin{eqnarray}
A_i^{\mp}  =  \frac{1}{\sqrt{2}} \sqrt{1 \! \mp \! \frac{J_{1}(\sigma_{1,i}\! +\! \sigma_{1,i+1}\! -\! \sigma_{2,i})}{\sqrt{J^{2}_{1}(\sigma_{1,i}\! +\! \sigma_{1,i+1}\! -\! \sigma_{2,i})^{2}\! +\! (J\Delta)^{2}}}}.
\label{AMP}
\end{eqnarray}
With the use of the eigenvalues \eqref{IsiHeiEigenVal} one can further rewrite the partition function \eqref{IsiHeiPartRE} into the following simplified form:
\begin{eqnarray}
Z =\sum_{\{ \sigma_{1,i} \} } \prod_{i=1}^{N} {\rm \textbf{T}}(\sigma_{1,i},\sigma_{1,i+1}),
\label{IsiHeiPartRERE}
\end{eqnarray}
where the effective Boltzmann’s factor ${\rm \textbf{T}}(\sigma_{1,i},\sigma_{1,i+1})$ is defined as follows:
\begin{eqnarray}
{\rm \textbf{T}}(\sigma_{1,i},\sigma_{1,i+1}) \!\!\!&=&\!\!\! 
\exp\left[\frac{\beta h_{I}}{2}(\sigma_{1,i} + \sigma_{1,i+\! 1})\right] 
\sum_{\sigma_{2,i}} \exp\left(\beta h_{I} \sigma_{2,i}\right) \sum_{j=1}^{4} \exp\left(- \beta E_{j,i}\right)
\nonumber \\ 
\!\!\!&=&\!\!\! \exp\left[\frac{\beta h_{I}}{2}(\sigma_{1,i} \!+\! \sigma_{1,i+\!1})\right]\Biggl \{\! 
\exp\left[-\frac{\beta}{4}(J \!+\! J_{1})\right]\! \cosh \left[\beta\left(\frac{J_{1}}{2} (\sigma_{1,i} 
\!+\! \sigma_{1,i+\!1}\!)\!-\! h_{H} \!-\! \frac{h_{I}}{2}\!\right) \right]
\Biggr. \nonumber \\ \!\!\!&+&\!\!\! \Biggl.
 \exp\left[\frac{\beta}{4}(- J + J_{1})\right] \cosh\left[\beta\left(\frac{J_{1}}{2}(\sigma_{1,i} + \sigma_{1,i+1})- h_{H} + \frac{h_{I}}{2}\right) \right]
\Biggr. \nonumber \\ \!\!\!&+&\!\!\! \Biggl.
 \exp\left[\frac{\beta J}{4} + \frac{\beta h_{I}}{2}\right] \cosh \left[\frac{\beta}{2}\sqrt{\left(J_{1}(\sigma_{1,i} + \sigma_{1,i+1}) - \frac{J_{1}}{2} \right)^{2} +  (J \Delta)^{2}} \right]
\Biggr. \nonumber \\ \!\!\!&+&\!\!\! \Biggl. 
 \exp\left[\frac{\beta J}{4} - \frac{\beta h_{I}}{2}\right]\! \cosh \left[\frac{\beta}{2}\sqrt{\left(J_{1}(\sigma_{1,i} \!+\! \sigma_{1,i+1}) \!+\! \frac{J_{1}}{2} \right)^{2} \!\!+\!  (J \Delta)^{2}} \right]\Biggr\}.
\label{IsiHeiTransfer}
\end{eqnarray}
The effective Boltzmann’s weight \eqref{IsiHeiTransfer} depends only on degrees of freedom of two Ising spins $\sigma_{1,i}$ and $\sigma_{1,i+1}$ and it can be alternatively viewed as the transfer matrix:
\begin{eqnarray}
\!\!\!\!&&\!\!\!\!
\rm {\textbf{T}}(\sigma_{1,i}, \sigma_{1,i+1})=
\begin{pmatrix}
{\rm T}\left(+\frac{1}{2},+\frac{1}{2}\right) \!&\! {\rm T}\left(+\frac{1}{2},-\frac{1}{2}\right) \\
{\rm T}\left(-\frac{1}{2},+\frac{1}{2}\right) \!&\! {\rm T}\left(-\frac{1}{2},-\frac{1}{2}\right)
\end{pmatrix} =
\begin{pmatrix}
{\rm T}_{1} \!&\! {\rm T}_{3} \\
{\rm T}_{3} \!&\! {\rm T}_{2}
\end{pmatrix},
\nonumber \\ \!\!\!\!&&\!\!\!\!
\label{TransMatrix}
\end{eqnarray}
where ${\rm T}_{i}$ ($i=1,2,3$) denote three different elements of the transfer matrix obtained from Eq. (\ref{IsiHeiTransfer}) by considering all four available states of two Ising spins $\sigma_{1,i}$ and $\sigma_{1,i+1}$: 
\begin{eqnarray}
{\rm T}_{1} \!\!\!&=&\!\!\! 
2\exp \left(\frac{\beta h_{I}}{2}\right)\Biggl\{{\rm e}^{- \frac{\beta}{4}(J \!+\! J_{1})} 
\cosh \left[\beta\left(\frac{J_{1}}{2} \!-\! h_{H} \!-\! \frac{h_{I}}{2}\right)\right] 
+ {\rm e}^{\frac{\beta}{4}(-J + J_{1})} \cosh \left[\beta\left(\frac{J_{1}}{2} - h_{H} + \frac{h_{I}}{2}\right) \right]\Biggr. \nonumber \\ \!\!\!&+&\!\!\!  \Biggl. \exp \left(\frac{\beta J}{4} \right)\left[{\rm e}^{\frac{\beta h_{I}}{2}}\cosh \left(\frac{\beta}{2}\sqrt{\left(\frac{J_{1}}{2}\right)^{2}+(J\Delta)^{2}}\right)
+{\rm e}^{-\frac{\beta h_{I}}{2}}\cosh \left(\frac{\beta}{2}\sqrt{\left(\frac{3J_{1}}{2}\right)^{2}+(J\Delta)^{2}}\right)\right] \Biggr\},
\nonumber \\ 
{\rm T}_{2}\!\!\!&=&\!\!\! 
2\exp \left(-\frac{\beta h_{I}}{2}\right)\Biggl\{{\rm e}^{- \frac{\beta}{4}(J \!+\! J_{1})} 
\cosh \left[\beta\left(\frac{J_{1}}{2} \!+\! h_{H} \!+\! \frac{h_{I}}{2}\right) \right]
+{\rm e}^{\frac{\beta}{4}(-J + J_{1})} \cosh \left[\beta\left(-\frac{J_{1}}{2} - h_{H} + \frac{h_{I}}{2}\right) \right]\Biggr. \nonumber \\ \!\!\!&+&\!\!\! \Biggl. \exp \left(\frac{\beta J}{4} \right)
\left[{\rm e}^{\frac{\beta h_{I}}{2}}\cosh\left(\frac{\beta}{2}\sqrt{\left(\frac{3J_{1}}{2}\right)^{2}+(J\Delta)^{2}}\right) + {\rm e}^{-\frac{\beta h_{I}}{2}}\cosh \left(\frac{\beta}{2}\sqrt{\left(\frac{J_{1}}{2}\right)^{2}+(J\Delta)^{2}}\right)\right] \Biggr\},
\nonumber \\ 
{\rm T}_{3}\!\!\!&=&\!\!\! 2\Biggl\{ \exp \left[- \frac{\beta}{4}(J \!+\! J_{1}) \right] \cosh \left[\beta\left(h_{H} \!+\! \frac{h_{I}}{2}  \right)\right] + \exp \left[\frac{\beta}{4}(-J + J_{1}) \right] \cosh \left[\beta\left(-h_{H} + \frac{h_{I}}{2}\right)\right]\Biggr. \nonumber \\ \!\!\!&+&\!\!\! \Biggl. 2 \exp \left(\frac{\beta J}{4} \right)\cosh \left(\frac{\beta}{2}\sqrt{\left(\frac{J_{1}}{2}\right)^{2}+(J\Delta)^{2}}\right)
\cosh \left(\frac{\beta h_{I}}{2}\right) \Biggr\}.
\label{TmatrixElements}
\end{eqnarray}
In the spirit of the transfer-matrix method, the final expression for the partition function can be finally expressed in terms of the transfer-matrix eigenvalues:
\begin{eqnarray}
Z = \sum_{\{ \sigma_{1,i} \} } \prod_{i=1}^{N}{\rm \textbf{T}}(\sigma_{1,i}, \sigma_{1,i+1})=\sum_{\sigma_{1,1}=\pm\frac{1}{2}}{\rm \textbf{T}}^{N}(\sigma_{1,1},\sigma_{1,1})
={\rm Tr}~{\rm \textbf{T}}^{N} = \rm{\lambda}_{+}^{N} + \rm{\lambda}_{-}^{N}.
\label{PartFinal}
\end{eqnarray} 
The transfer matrix \eqref{TransMatrix} can be readily diagonalized in order to obtain the eigenvalues 
$\rm{\lambda}_{\pm}$ defined through the transfer-matrix elements (\ref{TmatrixElements}):
\begin{eqnarray}
\lambda_{\pm} = \frac{1}{2} \left[{\rm T}_{1} + {\rm T}_{2} \pm \sqrt{({\rm T}_{1} - {\rm T}_{2})^{2} + 4 {\rm T}_{3}^{2}} \right].
\label{TransEigenVal}
\end{eqnarray}
In the thermodynamic limit $N \to \infty$, the Gibbs free energy of the spin-1/2 Ising-Heisenberg branched chain per unit cell is given by the largest transfer-matrix eigenvalue:
\begin{eqnarray}
g = -k_{\rm B} T \lim_{N\to\infty} \frac{1}{N}  \ln Z= -k_{\rm B} T \ln \lambda_{+}.
\label{Gibbs}
\end{eqnarray}
From the Gibbs free energy \eqref{Gibbs} one can subsequently derive the local magnetization of the Heisenberg and Ising spins as well as the total magnetization (per spin) 
of the spin-1/2 Ising-Heisenberg branched chain:
\begin{eqnarray}
m_{H} = \frac{1}{2}\left\langle \hat{S}_{1,i}^{z} + \hat{S}_{2,i}^{z}\right\rangle = -\frac{1}{2}\left(\frac{\partial  g}{\partial h_{H}}\right)_{T}\!\!\!, \quad
m_{I} = \frac{1}{2}\left\langle \hat{\sigma}_{1,i}^{z} + \hat{\sigma}_{2,i}^{z}\right\rangle = -\frac{1}{2}\left(\frac{\partial g}{\partial h_{I}}\right)_{T}\!\!\!, \quad
m =\frac{m_{H} + m_{I}}{2}.
\label{HeiMag}
\end{eqnarray}
From the exact result \eqref{Gibbs} for the Gibbs free energy one can eventually derive other quantities such as magnetic susceptibility, entropy and specific heat using the standard relations of thermodynamics. Besides, one may also investigate a strength of the bipartite  entanglement inside of the Heisenberg dimers using the quantity concurrence \cite{woo98}, which can be calculated from the pair correlation functions and the local magnetization \cite{ami08,hor09}:
\begin{eqnarray}
\!\!\!\!&&\!\!\!\!
C = \! \max\! \left\{ 0, 4 \left|C^{xx} \right| \!-\!  2 \sqrt{\left(\frac{1}{4} \!+\! C^{zz}\right)^{2} \!- \left(m_{H}\right)^{2}}\right\}. \nonumber \\ \!\!\!\!&&\!\!\!\!
\label{Concurrence}
\end{eqnarray}
Here, $C^{xx}= \left\langle \hat{S}_{1,i}^{x} \hat{S}_{2,i}^{x}\right\rangle$ and $C^{zz}= \left\langle \hat{S}_{1,i}^{z} \hat{S}_{2,i}^{z}\right\rangle$ denote two spatial components of the pair correlation function between the nearest-neighbor Heisenberg spins and $m_{H}$ is the respective local magnetization given by Eq. \eqref{HeiMag}. The pair correlation functions can be straightforwardly derived from the Gibbs free energy \eqref{Gibbs} according to the relations:
\begin{eqnarray}
C^{xx} = -\frac{1}{2N} \frac{\partial \ln Z}{\partial (\beta J \Delta)}, \quad C^{zz} = -\frac{1}{N} \frac{\partial \ln Z}{\partial (\beta J)},
\label{pairFc}
\end{eqnarray}
but their explicit form is too cumbersome to write it down here explicitly.

\section{Theoretical results}
Let us proceed to a discussion of the most interesting results obtained for the spin-1/2 Ising-Heisenberg branched chain by assuming the isotropic antiferromagnetic Heisenberg coupling constant $J>0$, while keeping the anisotropy parameter fixed at the constant value of 
$\Delta=1$ for illustrative purposes. Throughout this section, the coupling constant $J>0$ will serve as an energy unit when defining a relative strength of the magnetic field $h/J$, the dimensionless temperature $k_{\rm B}T/J$ and the interaction ratio $J_{1}/J$. To reduce the number of free parameters, the Land\'e $g$-factors of the Heisenberg and Ising spins assigned to ${\rm Cu}^{2+}$ and ${\rm Fe}^{3+}$ magnetic ions were set equal to each other, thereby ensuring that the 'local' magnetic fields exert an equal strength $h \equiv h_{H}=h_{I}$. 
\begin{figure}[t]
\centering
\hspace{-0.0cm}
\includegraphics[width=0.15\textwidth]{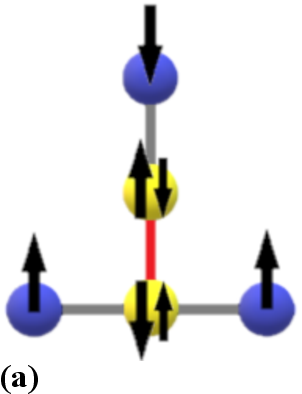}
\hspace{0.8cm}
\includegraphics[width=0.15\textwidth]{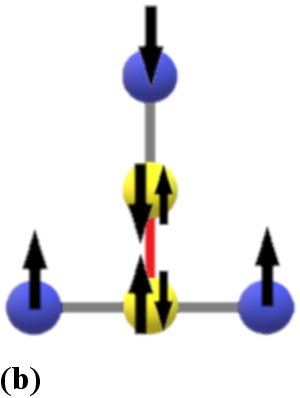}
\hspace{0.8cm}
\includegraphics[width=0.15\textwidth]{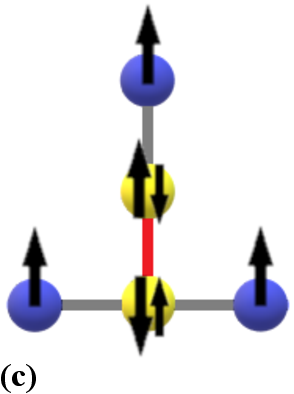}
\hspace{0.8cm}
\includegraphics[width=0.15\textwidth]{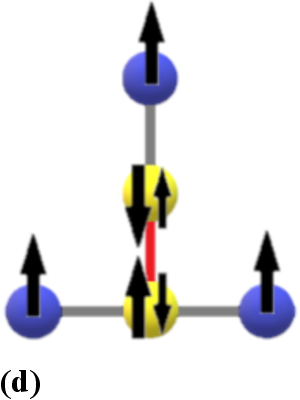}
\hspace{0.8cm}
\includegraphics[width=0.15\textwidth]{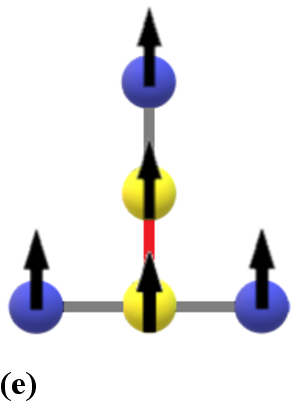}
\hspace{0.0cm}
\vspace{-0.3cm}
\caption{A schematic illustration of typical configurations realized within all possible ground states of the spin-1/2 Ising-Heisenberg branched chain: (a)-(b) the quantum antiferromagnetic phases $\left|I\right\rangle$ and $\left|I'\right\rangle$; (c)-(d) the quantum ferrimagnetic phases $\left|II\right\rangle$ and $\left|II'\right\rangle$; (e) the classical ferromagnetic phase $\left|III\right\rangle$. The pair of arrows situated at the Heisenberg spins symbolizes a singlet-like quantum superposition, whereby the size of arrows reflects different magnitudes of the respective probability amplitudes (\ref{ampStates}).}
\label{state}
\end{figure}

\begin{figure*}[t]
\centering
\vspace{-1.1cm}
\includegraphics[width=0.95\textwidth]{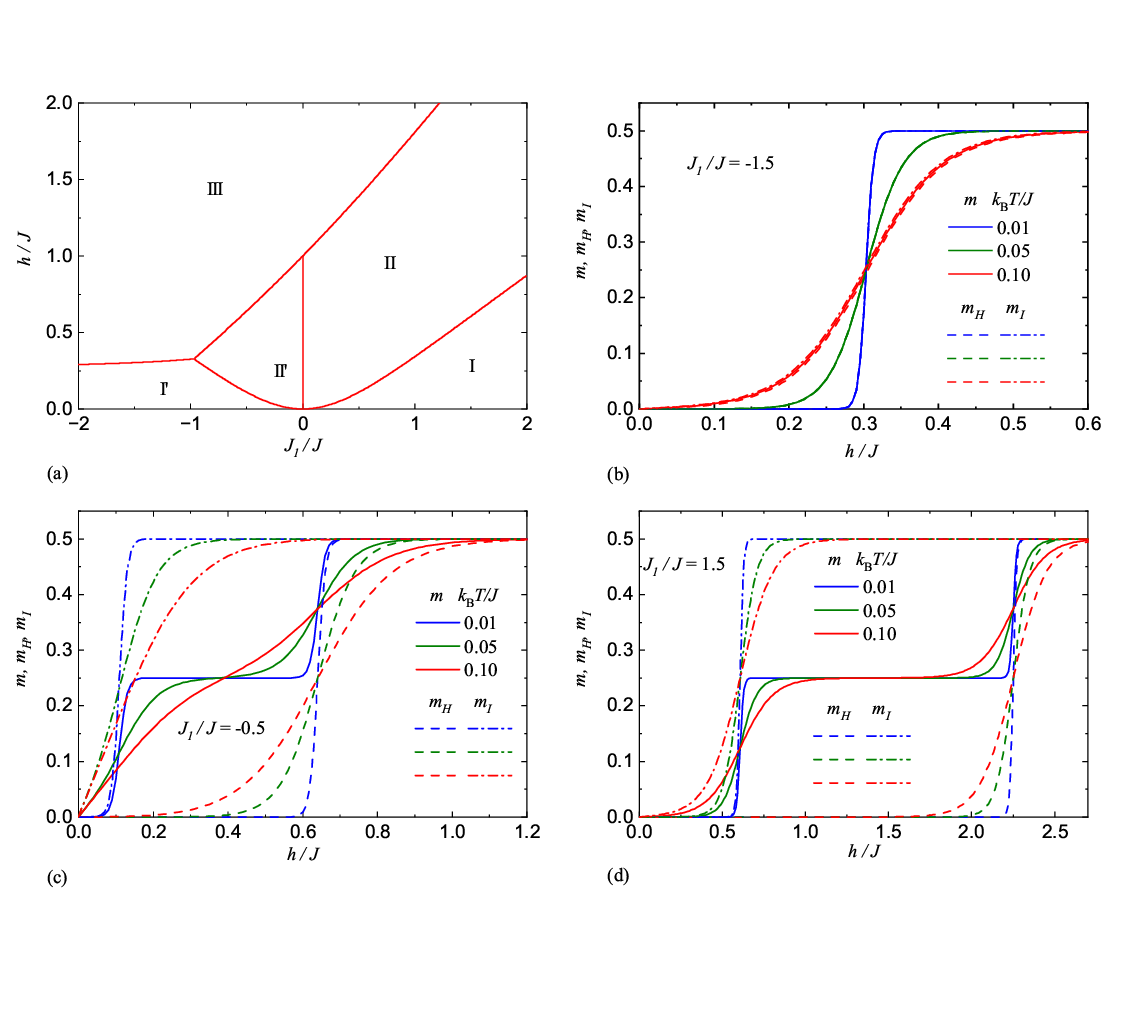}
\vspace{-2cm}
\caption{(a) The ground-state phase diagram for the spin-1/2 Ising-Heisenberg branched chain in the $J_1/J-h/J$ plane by assuming the isotropic value of the parameter $\Delta=1$. The regions I(I'), II(II') and III correspond to a quantum antiferromagnetic, a quantum ferrimagnetic and a classical ferromagnetic phase, respectively; (b)-(d) Magnetic-field dependencies of the local and total magnetization given by Eq. \eqref{HeiMag} for the spin-1/2 Ising-Heisenberg branched chain at three different temperatures and three selected values of the interaction ratio: (b) $J_{1}/J=-1.5$; (c) $J_{1}/J=-0.5$; (d) $J_{1}/J=1.5$.}
\label{phase}
\end{figure*}

By comparing energies of all available eigenstates of the spin-1/2 Ising-Heisenberg branched chain one finds just three different ground states schematically shown in Fig. \ref{state} and mathematically given by the eigenvectors:
\begin{eqnarray}
\left|I, I'\right\rangle=\prod_{i=1}^{N}\left(a_{-}\left|\uparrow\right\rangle_{S_{1,i}}\left|\downarrow\right\rangle_{S_{2,i}}-a_{+}\left|\downarrow\right\rangle_{S_{1,i}}\left|\uparrow\right\rangle_{S_{2,i}}\right)\left|\uparrow\right\rangle_{\sigma_{1,i}}\left|\downarrow\right\rangle_{\sigma_{2,i}},
\label{qaf}
\end{eqnarray}
\begin{eqnarray}
\left|II, II'\right\rangle=\prod_{i=1}^{N}\left(b_{-}\left|\uparrow\right\rangle_{S_{1,i}}\left|\downarrow\right\rangle_{S_{2,i}}-b_{+}\left|\downarrow\right\rangle_{S_{1,i}}\left|\uparrow\right\rangle_{S_{2,i}}\right)\left|\uparrow\right\rangle_{\sigma_{1,i}}\left|\uparrow\right\rangle_{\sigma_{2,i}},
\label{qfi}
\end{eqnarray}
\begin{eqnarray}
\left|III\right\rangle=\prod_{i=1}^{N}\left|\uparrow\right\rangle_{S_{1,i}}\left|\uparrow\right\rangle_{S_{2,i}}\left|\uparrow\right\rangle_{\sigma_{1,i}}\left|\uparrow\right\rangle_{\sigma_{2,i}}.
\label{cfo}
\end{eqnarray}
While the factorized form of the third ground state (\ref{cfo}) is consistent with the classical nature of the fully polarized ferromagnetic state $\left|III\right\rangle$, the quantum antiferromagnetic phase $\left|I, I'\right\rangle$ and the quantum ferrimagnetic phase $\left|II, II'\right\rangle$ described by Eqs. \eqref{qaf} and \eqref{qfi} exhibit bipartite quantum entanglement among the Heisenberg spin pairs $S_{1,i}$-$S_{2,i}$ unambiguously characterized through the probability amplitudes $a_{\pm}$ and $b_{\pm}$:
\begin{eqnarray}
a_{\pm}= \frac{1}{\sqrt{2}} \sqrt{1 \pm \frac{\frac{3}{2} \frac{J_{1}}{J}}{\sqrt{\left(\frac{3}{2} \frac{J_{1}}{J}\right)^{2}+\Delta^{2}}}},
\qquad
b_{\pm}= \frac{1}{\sqrt{2}} \sqrt{1 \pm \frac{\frac{1}{2} \frac{J_{1}}{J}}{\sqrt{\left(\frac{1}{2} \frac{J_{1}}{J}\right)^{2}+\Delta^{2}}}}.
\label{ampStates}
\end{eqnarray} 
From this perspective, it would be insightful to compare the strength of the bipartite quantum entanglement emerging within the quantum antiferromagnetic \eqref{qaf} and ferrimagnetic \eqref{qfi} phases using the concept of the concurrence \eqref{Concurrence},
which acquires within these two ground states the following values:
\begin{eqnarray}
C_{\left|I, I'\right\rangle} = \frac{\Delta}{\sqrt{\left(\frac{3}{2} \frac{J_{1}}{J}\right)^{2}+\Delta^{2}}},
\qquad
C_{\left|II, II'\right\rangle} = \frac{\Delta}{\sqrt{\left(\frac{1}{2} \frac{J_{1}}{J}\right)^{2}+\Delta^{2}}}.
\label{zeroCon}
\end{eqnarray}
It is obvious from Eq. \eqref{zeroCon} that the bipartite quantum entanglement of the Heisenberg spin pairs is consistently stronger within the quantum ferrimagnetic phase $\left|II, II'\right\rangle$ compared to the quantum antiferromagnetic phase $\left|I, I'\right\rangle$ for a given set of the interaction ratio $J_1/J$ and the exchange anisotropy $\Delta$, because a fully polarized nature of the enclosing Ising spins tends to bring a singlet-like state closer to a perfect singlet-dimer state as one of the Bell states.  

For the sake of completeness, let us also quote eigenenergies (per unit cell) corresponding to the quantum antiferromagnetic phase $\left|I, I'\right\rangle$ 
\begin{eqnarray}
E_{I/I'}=-\frac{J}{4}-\frac{1}{2}\sqrt{\left(\frac{3J_{1}}{2}\right)^{2}+(J\Delta)^{2}},
\label{engAB}
\end{eqnarray}
the quantum ferrimagnetic phase $\left|II, II'\right\rangle$ 
\begin{eqnarray}
E_{II/II'}=-\frac{J}{4}-\frac{1}{2}\sqrt{\left(\frac{J_{1}}{2}\right)^{2}+(J\Delta)^{2}}-h,
\label{engCD}
\end{eqnarray}
 the classical ferromagnetic phase $\left|III\right\rangle$ 
\begin{eqnarray}
E_{III}=\frac{J}{4}+\frac{3J_{1}}{4}-2h.
\label{engE}
\end{eqnarray}
It should be stressed that the ground states $\left|I\right\rangle$ and $\left|I'\right\rangle$ have the same energy and they just differ by a relative orientation of the singlet-like state of the Heisenberg spin pairs with respect to their neighboring Ising spins [see Fig. \ref{state}(a)-(b)] depending on whether the Ising coupling constant is antiferromagnetic $J_{1}>0$ or ferromagnetic $J_{1}<0$. The same statement holds true also for the quantum ferrimagnetic phases $\left|II\right\rangle$ and $\left|II'\right\rangle$ schematically drawn in Fig. \ref{state}(c)-(d), respectively. A direct comparison of all three ground-state energies (\ref{engAB})-(\ref{engE}) affords exact expressions for the relevant ground-state phase boundaries:

1. phase boundary $\left|I\right\rangle$/$\left|II\right\rangle$ and $\left|I'\right\rangle$/$\left|II'\right\rangle$:
\begin{eqnarray}
\frac{h}{J}=\frac{1}{2}\left[\sqrt{\left(\frac{3}{2}\frac{J_{1}}{J}\right)^{2}+\Delta^{2}}-\sqrt{\left(\frac{1}{2}\frac{J_{1}}{J}\right)^{2}+\Delta^{2}}\right],
\label{FirstB}
\end{eqnarray} 

2. phase boundary $\left|I'\right\rangle$/$\left|III\right\rangle$:
\begin{eqnarray}
\frac{h}{J}=\frac{1}{4}+\frac{3}{8}\frac{J_{1}}{J}+\frac{1}{4}\sqrt{\left(\frac{3}{2}\frac{J_{1}}{J}\right)^{2}+\Delta^{2}},
\label{SecondB}
\end{eqnarray}

3. phase boundary $\left|II\right\rangle$/$\left|III\right\rangle$ and $\left|II'\right\rangle$/$\left|III\right\rangle$:
\begin{eqnarray}
\frac{h}{J}=\frac{1}{2}+\frac{3}{4}\frac{J_{1}}{J}+\frac{1}{2}\sqrt{\left(\frac{1}{2}\frac{J_{1}}{J}\right)^{2}+\Delta^{2}},
\label{Third}
\end{eqnarray}

4. phase boundary $\left|II\right\rangle$/$\left|II'\right\rangle$:
\begin{eqnarray}
J_{1}/J=0.
\label{ThiB}
\end{eqnarray}

Using these exact expressions, we have plotted  in Fig. \ref{phase}(a) the overall ground-state phase diagram of the spin-1/2 Ising-Heisenberg branched chain in the $J_1/J-h/J$ plane. The parameter regions denoted in Fig. \ref{phase}(a) as I(I'), II(II') and III correspond to the quantum antiferromagnetic (QAF), the quantum ferrimagnetic (QFI) and the classical ferromagnetic (CF) phase, respectively. Note furthermore that the coexistence lines (\ref{FirstB})-(\ref{Third}) correspond to discontinuous magnetic-field-driven phase transitions. All three phases QAF, QFI and CF coexist together in Fig. \ref{phase}(a) at a triple coexistence point emerging for the specific value of the interaction ratio $J_1/J \approx -0.97$. Below this value, the QFI ground state $\left|II'\right\rangle$ is absent and only the two ground states $\left|I'\right\rangle$ and $\left|III\right\rangle$ appear. 

To verify this observation, we present magnetic-field dependencies of the local and total magnetization in Fig. \ref{phase}(b) for the interaction ratio $J_{1}/J=-1.5$ and three distinct temperature values. Under this condition, the local magnetization $m_{H}$ and $m_{I}$ defined as a single-site  magnetization of the Heisenberg and Ising spins are zero until the saturation field $h_{s}/J\approx 0.3$ is reached. This consequently leads to a zero magnetization plateau of the total magnetization $m$, which is consistent with the antiferromagnetic alignment of the Ising spins and the singlet-like ordering of the Heisenberg spin pairs realized within the QAF ground state $\left|I'\right\rangle$ [see Fig. \ref{state}(b)]. The observed single magnetization jump can thus be attributed to a discontinuous magnetic-field-driven phase transition from the QAF ground state $\left|I'\right\rangle$ to the CF ground state $\left|III\right\rangle$. It should be also noted that the true magnetization jump exists at zero temperature only, which however changes to a steep but continuous rise upon increasing of temperature.

Next, the local and total magnetization of the spin-1/2 Ising-Heisenberg branched chain are plotted against the magnetic field in Fig. \ref{phase}(c) for another value of the interaction ratio $J_{1}/J=-0.5$ and three distinct temperatures. The spin-1/2 Ising-Heisenberg branched chain undergoes in this particular case two magnetic-field-driven phase transitions: initially from the QAF ground state $\left|I'\right\rangle$ towards the QFI ground state $\left|II'\right\rangle$ subsequently followed by a transition from the QFI ground state $\left|II'\right\rangle$ towards the CF ground state $\left|III\right\rangle$. In agreement with this statement, the local magnetization of the Ising spins $m_{I}$ does not contribute to the total magnetization $m$ only at sufficiently low magnetic fields $h/J\lesssim 0.11$ stabilizing the QAF ground state $\left|I'\right\rangle$ and then it undergoes sudden rise towards the saturated value acquired within the QFI ground state $\left|II'\right\rangle$ and the CF  ground state $\left|III\right\rangle$. On the contrary, the local magnetization of the Heisenberg spins $m_{H}$ does not contribute to the total magnetization $m$ unless the magnetic field exceeds its saturation value $h_{s}/J\approx 0.64$. 

Finally, we illustrate in Fig. \ref{phase}(d) the local and total magnetization of the spin-1/2 Ising-Heisenberg branched chain for a fixed value of the interaction ratio $J_{1}/J=1.5$. The behavior observed here closely resembles that of the previous cases, with two intermediate magnetization plateaus and two magnetic-field-induced transitions: first from the QAF phase $\left|I\right\rangle$ towards the QFI phase $\left|II\right\rangle$, followed by a transition from the QFI phase $\left|II\right\rangle$ towards the CF phase $\left|III\right\rangle$. The primary quantitative difference lies in a somewhat wider intermediate one-half magnetization plateau corresponding to the QFI phase $\left|II\right\rangle$, which generally broadens with an increase in the interaction ratio $J_{1}/J$.

\begin{figure}[t]
\centering
\vspace{-0.2cm}
\includegraphics[width=0.45\textwidth]{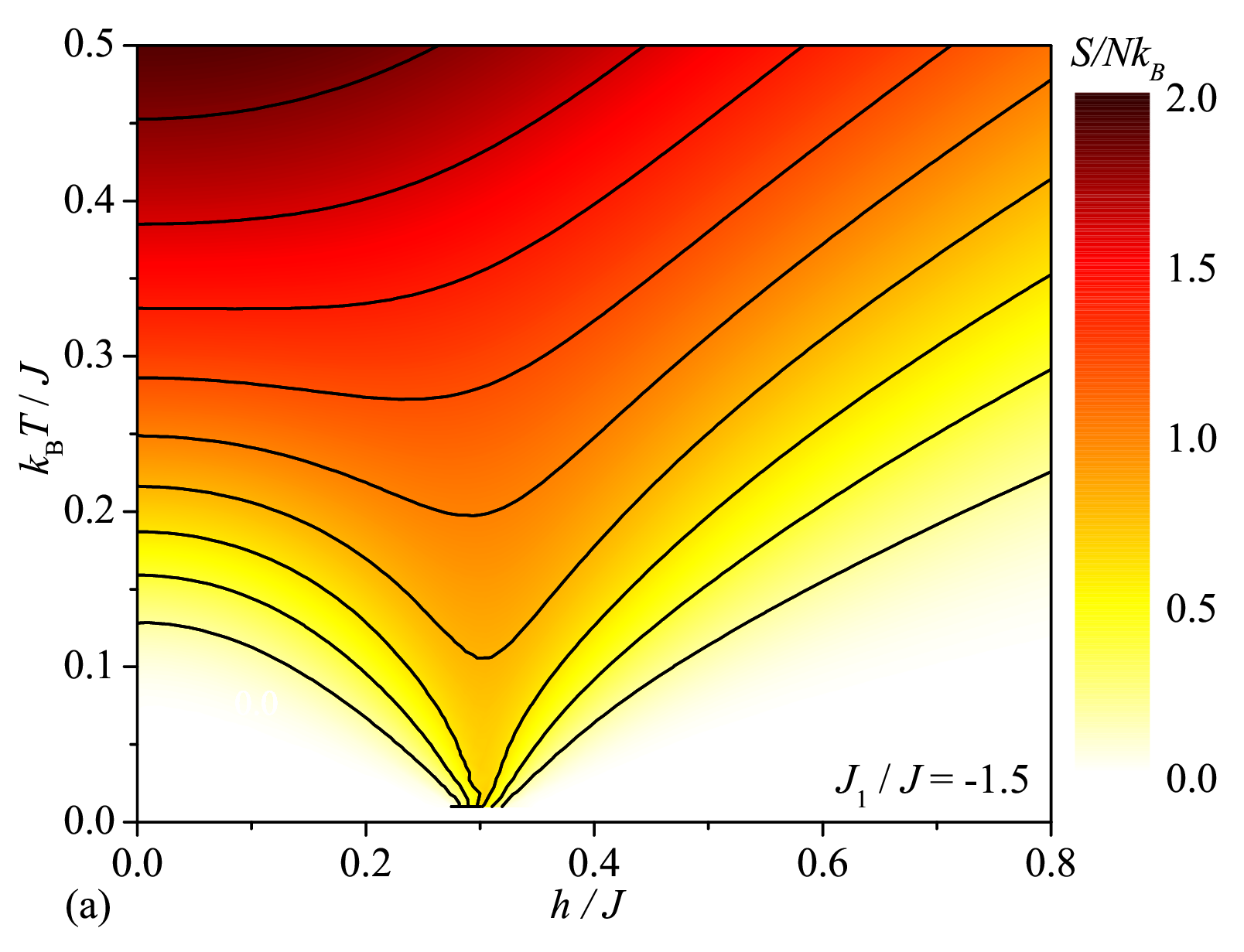}
\hspace{0.5cm}
\includegraphics[width=0.45\textwidth]{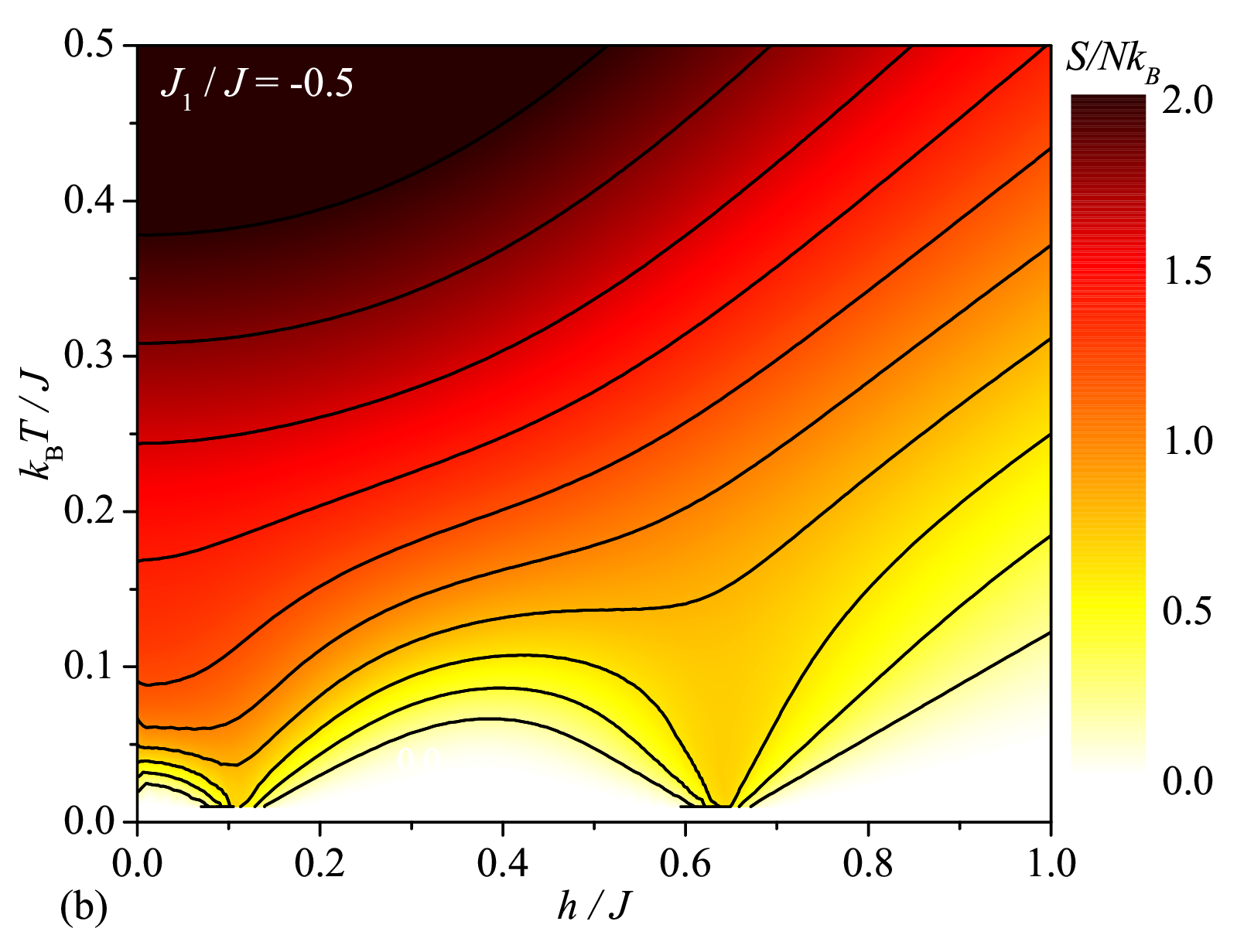}
\includegraphics[width=0.45\textwidth]{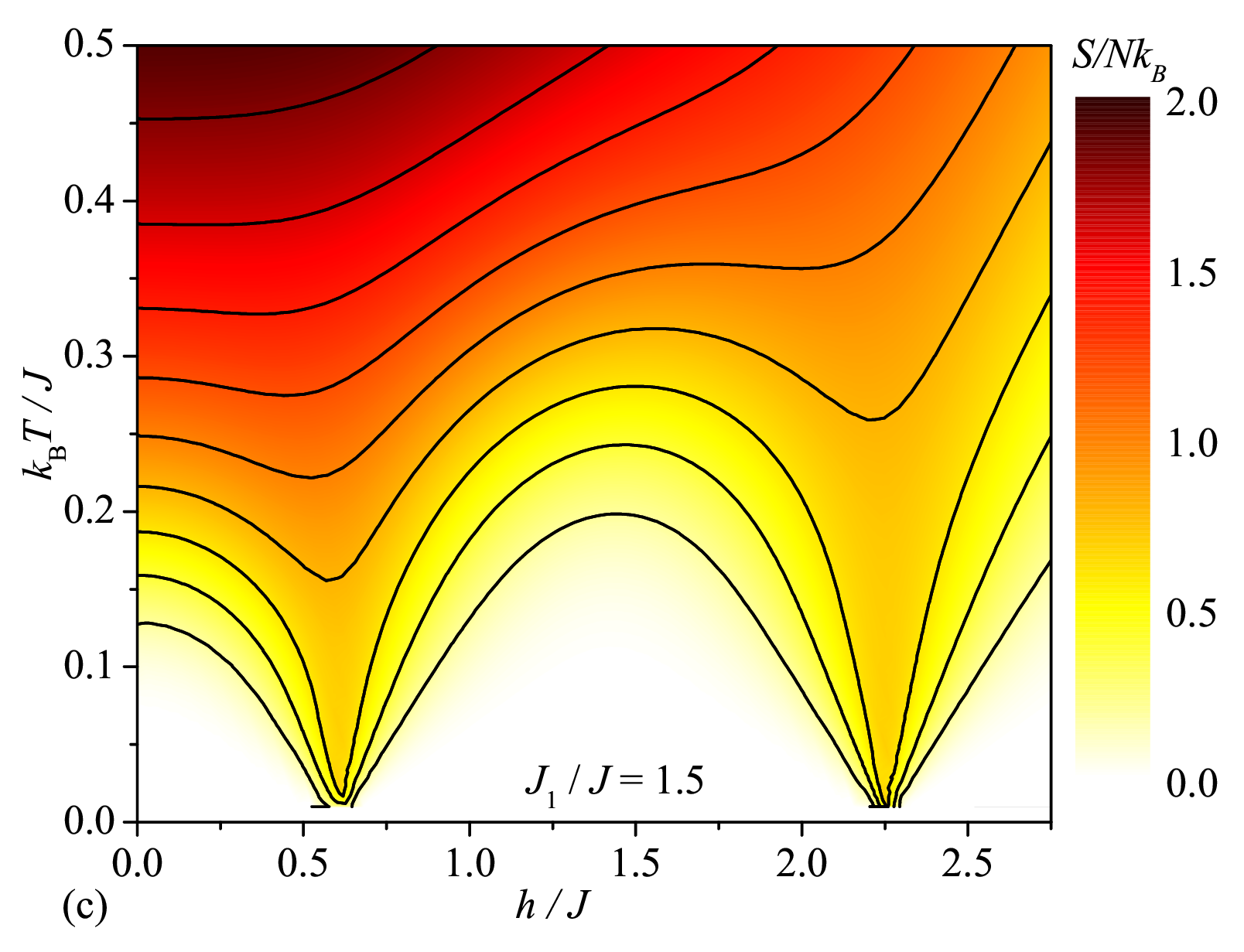}
\vspace{-0.2cm}
\caption{A density plot of the reduced entropy $S/N k_{\rm B}$ of the spin-1/2 Ising-Heisenberg branched chain in the magnetic field versus temperature plane $h/J-k_{\rm B}T/J$ for three selected values of the interaction ratio: (a) $J_{1}/J = -1.5$, (b) $J_{1}/J = -0.5$, and (c) $J_{1}/J = 1.5$.}
\label{mce}
\end{figure}

An independent validation of the aforementioned magnetization scenario is provided through a comprehensive analysis of the magnetic entropy, as depicted in Fig. \ref{mce} in the form of a density plot in the magnetic field versus temperature plane ($h/J$ - $k_{\rm B}T/J$). The isoentropy lines shown in Fig. \ref{mce} bring further insight into the adiabatic demagnetization process of the spin-1/2 Ising-Heisenberg branched chain. It is evident from Fig. \ref{mce} that a sharp decrease (or increase) in temperature is observed upon decreasing (or increasing) the magnetic field above (below) each magnetic-field-driven phase transition. Consistent with this observation, the isoentropy lines tend towards absolute zero temperature at a single magnetic field for the specific case $J_1/J = -1.5$, owing to the absence of the quantum ferrimagnetic phase $|II'\rangle$ at moderate magnetic fields [refer to Fig. \ref{mce}(a)]. Conversely, for the other two specific cases, $J_1/J = -0.5$ and $1.5$, the isoentropy lines approach absolute zero temperature at two distinct magnetic fields due to the presence of the quantum ferrimagnetic phase $|II'\rangle$ at moderate magnetic fields.

\begin{figure}[t]
\centering
\vspace{-0.5cm}
\includegraphics[width=0.5\textwidth]{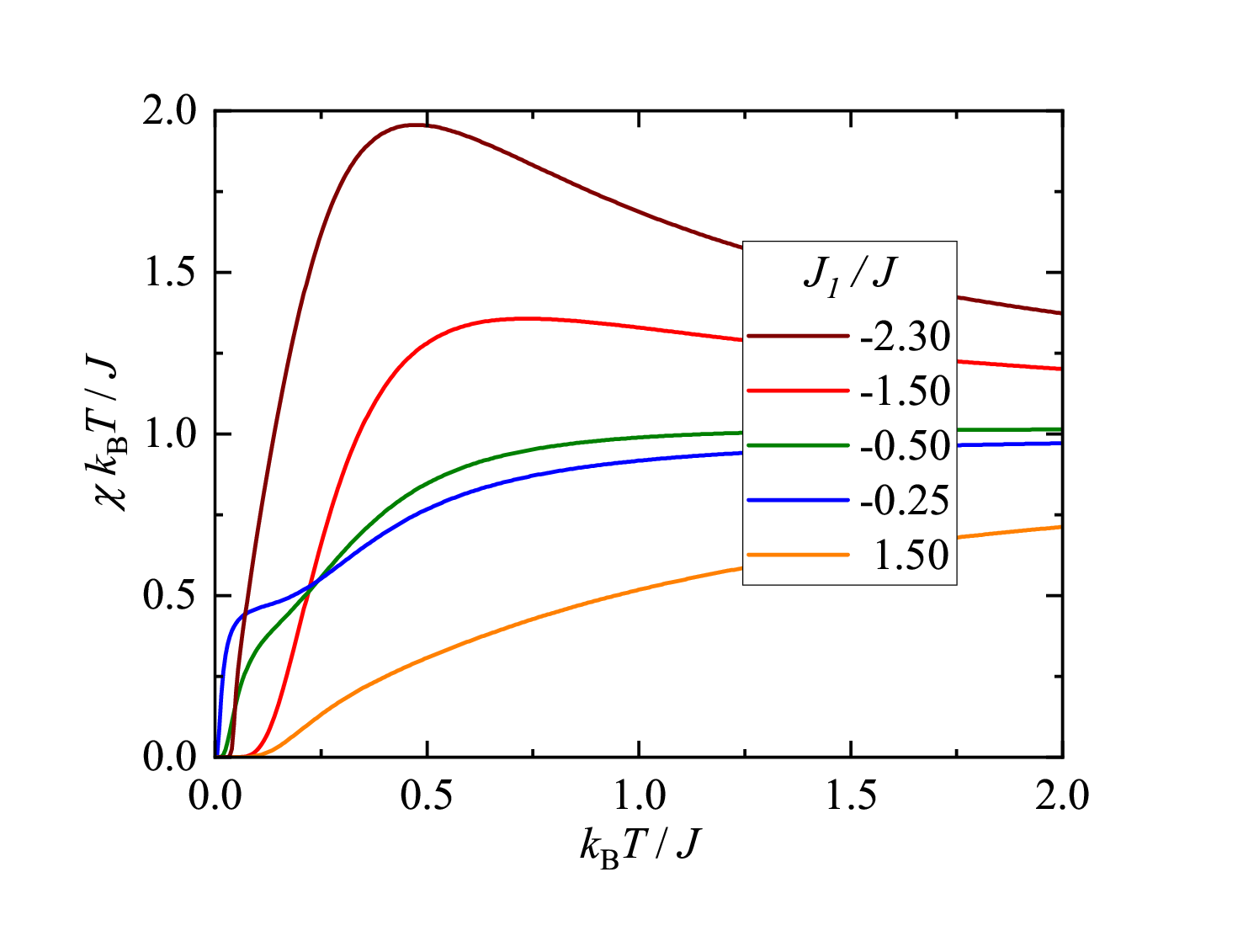}
\vspace{-0.9cm}
\caption{Temperature dependence of the initial magnetic susceptibility times the temperature $\chi k_{\rm B}T/J$ of the spin-1/2 Ising-Heisenberg branched chain for a few different values of the interaction ratio $J_{1}/J$.}
\label{teorXT}
\end{figure}

The temperature dependence of the initial magnetic susceptibility times temperature product is illustrated in Fig. \ref{teorXT} for various values of the interaction ratio $J_{1}/J$. It is noteworthy that the spin-1/2 Ising-Heisenberg branched chain exhibits the quantum antiferromagnetic (QAF) ground state in the zero-field limit for any value of the interaction ratio $J_{1}/J$ [refer to Fig. \ref{phase}(a)]. This aligns with the vanishing behavior of the susceptibility times temperature product as temperature approaches zero. It is apparent from Fig. \ref{teorXT} that the product of susceptibility and temperature gradually tends to zero as temperature decreases for the interaction ratio $J_{1}/J = 1.5$. For the interaction ratio $J_{1}/J = -0.25$, the product $\chi T$ contrarily initially declines, followed by maintaining a nearly constant value below the temperature $k_{\rm B}T/J \lesssim 0.25$, before rapidly decreasing to zero as absolute zero temperature is approached. The notable temperature dependence with a nearly constant value of the susceptibility times temperature product observed at moderate temperatures is attributed to the ferromagnetic nature of the Ising coupling constant $J_{1}<0$, which has opposite tendency of increasing of the product $\chi T$ upon decreasing of temperature. Indeed, the reinforcement of the ferromagnetic coupling constant $J_{1}$ leads to an increase in the susceptibility times temperature product as temperature decreases, as evidenced in Fig. \ref{teorXT} for specific values of the interaction ratio $J_{1}/J = -1.5$ and $-2.3$, where the product $\chi T$ initially reaches a rounded maximum before ultimately approaching zero as temperature decreases.

\begin{figure*}[t]
\centering
\vspace{-0.4cm}
\includegraphics[width=1.0\textwidth]{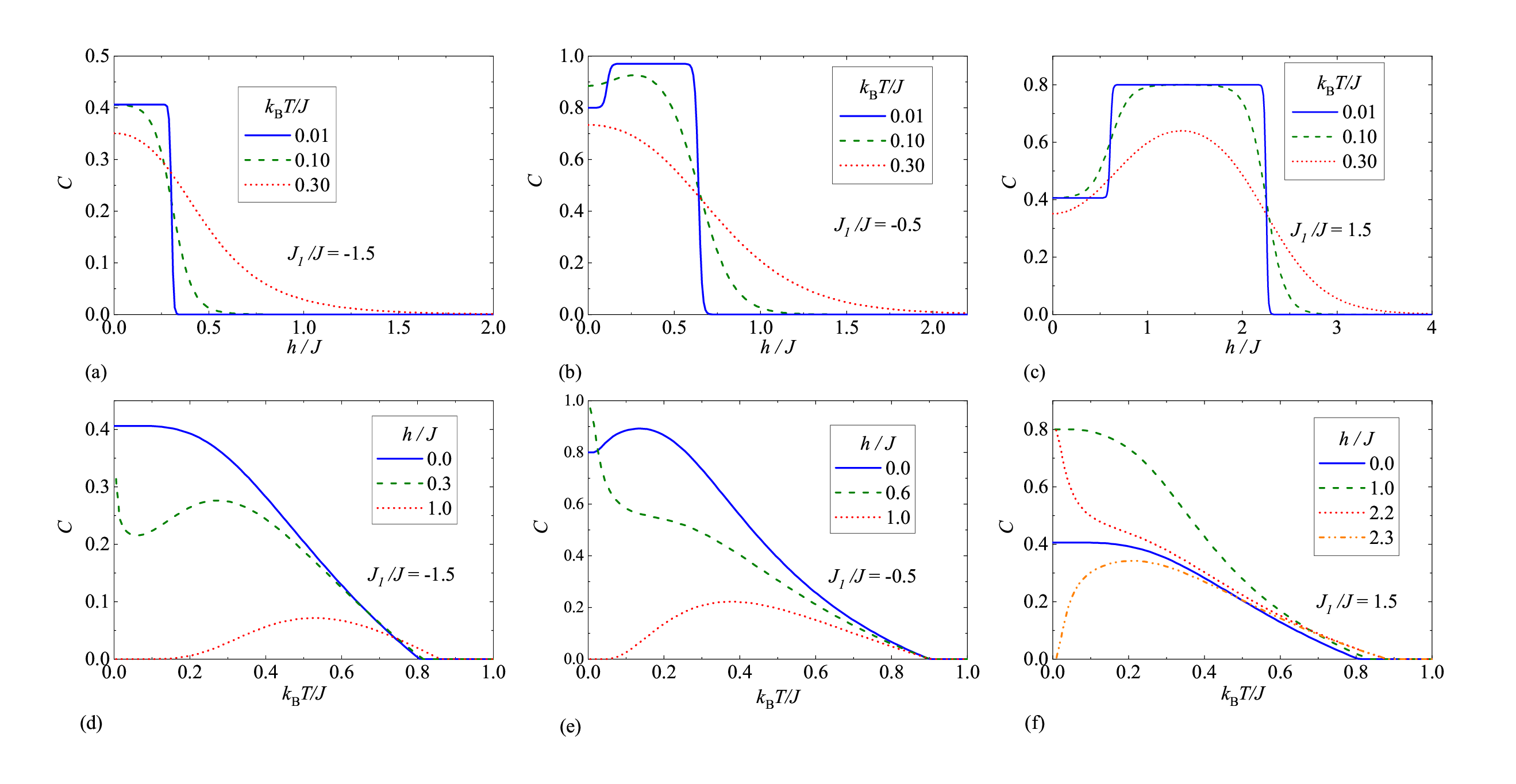} 
\vspace{-1.3cm}
\caption{Magnetic-field (a)-(c) and temperature (d)-(f) dependencies of the concurrence \eqref{Concurrence} calculated for the Heisenberg spin pairs of the spin-1/2 Ising-Heisenberg branched chain at three different values of the interaction ratio: (a) and (d) $J_{1}/J=-1.5$; 
(b) and (e) $J_{1}/J=-0.5$; (c) and (f) $J_{1}/J=1.5$.}
\label{con}
\end{figure*}

To better understand the quantum nature of the spin-1/2 Ising-Heisenberg branched chain, we conduct a detailed analysis of the concurrence \eqref{Concurrence}, which quantifies the strength of bipartite entanglement within the Heisenberg spin pairs. In Fig. \ref{con}(a), we illustrate the magnetic-field dependence of the concurrence for a fixed interaction ratio $J_{1}/J=-1.5$ and three distinct temperatures. At sufficiently low magnetic fields, the concurrence is non-zero due to the presence of the quantum antiferromagnetic (QAF) ground state $\left|I'\right\rangle$, characterized by singlet-like states of the Heisenberg dimers. This state breaks down at the magnetic-field-induced phase transition $h_{c}/J\approx 0.3$, leading to the emergence of the fully polarized classical ferromagnetic (CF) ground state $\left|III\right\rangle$. Fig. \ref{con}(a) demonstrates that increasing temperature generally smears out the stepwise magnetic-field dependence of the concurrence. The corresponding thermal variations of the concurrence are depicted in Fig. \ref{con}(d) for three selected magnetic field values. At zero magnetic field, the concurrence exhibits a standard monotonous decline with increasing temperature until it completely vanishes at a  threshold temperature. However, a more intricate temperature dependence is observed for the magnetic field $h/J=0.3$, around which the magnetic-field-driven phase transition occurs at zero temperature. Under this condition, the concurrence undergoes an abrupt decrease to a local minimum at the lowest temperatures, followed by a gradual rise to a local maximum at moderate temperatures, before eventually tending to zero with further temperature increase. Moreover, the concurrence initially starts from zero due to the CF ground state $\left|III\right\rangle$ at the magnetic field $h/J=1$, which exceeds the saturation value. However, the concurrence exhibits a striking double-reentrant behavior, developing first at a lower threshold temperature $k_{\rm{B}}T/J\approx 0.2$, then reaching a rounded maximum at a moderate temperature, before returning to zero at an upper threshold temperature $k_{\rm{B}}T/J\approx 0.85$. This double-reentrant behavior is attributed to the thermal population of the QAF phase $\left|I'\right\rangle$.

Next, we depict in Fig. \ref{con}(b) the concurrence as a function of the magnetic field for a fixed interaction ratio $J_{1}/J=-0.5$ and three distinct temperatures. In contrast to the previous case, the spin-1/2 Ising-Heisenberg branched chain encompasses at zero temperature all three possible ground states: QAF, QFI, and CF. Consequently, the concurrence initially attains a relatively large value ($C \approx 0.8$) due to the QAF ground state $\left|I'\right\rangle$ realized at sufficiently low magnetic fields ($h/J< 0.11$). This is followed by an almost maximal value ($C \approx 0.97$) attributed to the QFI ground state $\left|II'\right\rangle$, before the concurrence abruptly drops to zero near the saturation field ($h_{s}/J\approx 0.64$) at very low temperature ($k_{\rm{B}}T/J=0.01$), as observed in Fig. \ref{con}(b).
Furthermore, Fig. \ref{con}(b) reveals that the concurrence may become non-zero above the saturation field ($h_{s}/J\approx 0.64$) at slightly higher temperatures ($k_{\rm{B}}T/J=0.1$ and 0.3) due to thermal excitations from the fully polarized CF ground state $\left|III\right\rangle$ to the quantum states $\left|II'\right\rangle$ and $\left|I'\right\rangle$. To gain a comprehensive understanding, Fig. \ref{con}(e) illustrates the concurrence against temperature for the same value of the interaction ratio $J_{1}/J=-0.5$ and three different magnetic field values. At zero magnetic field, the concurrence starts from a finite value ($C\approx0.8$) corresponding to the QAF ground state, then steadily rises to reach a rounded maximum before gradually tending to zero. The prominent rounded peak of the concurrence can be attributed to thermal excitations from the less entangled QAF phase to the QFI phase with stronger entanglement. If the magnetic field $h/J=0.6$ favors the QFI ground state, the concurrence sharply drops from its almost maximal value within a narrow temperature range ($k_{\rm{B}}T/J\in(0;0.08)$) and then decreases steadily before showing a more abrupt thermally-induced decline. At a higher magnetic field ($h/J = 1$) exceeding the saturation value, the concurrence exhibits a double-reentrant behavior quite analogous to the previously described case for $J_{1}/J=-1.5$.

Last but not least, we have depicted in Fig. \ref{con}(c) and (f) the magnetic-field and temperature dependencies of the concurrence for the fixed value of the interaction ratio $J_{1}/J=1.5$. It is evident that the results shown in Fig. \ref{con}(c) and (f) are qualitatively very similar to those displayed in Fig. \ref{con}(b) and (e) for the other specific case $J_{1}/J=-0.5$. The main quantitative difference lies in the strength of bipartite entanglement, i.e., the magnitude of concurrence, which generally weakens with an increase in the absolute value of the interaction ratio $|J_{1}|/J$. Notably, the concurrence attains the same zero-temperature asymptotic value within the QAF ground states $\left|I\right\rangle$ and $\left|I'\right\rangle$ for two specific values of the interaction ratio $J_{1}/J=1.5$ and $-1.5$, respectively.

\section{Magnetic data of polymeric compound Fe$_2$ Cu$_2$: \\ theory vs. experiment}

\begin{figure*}[t]
\centering
\vspace{-0.5cm}
\includegraphics[width=0.99\textwidth]{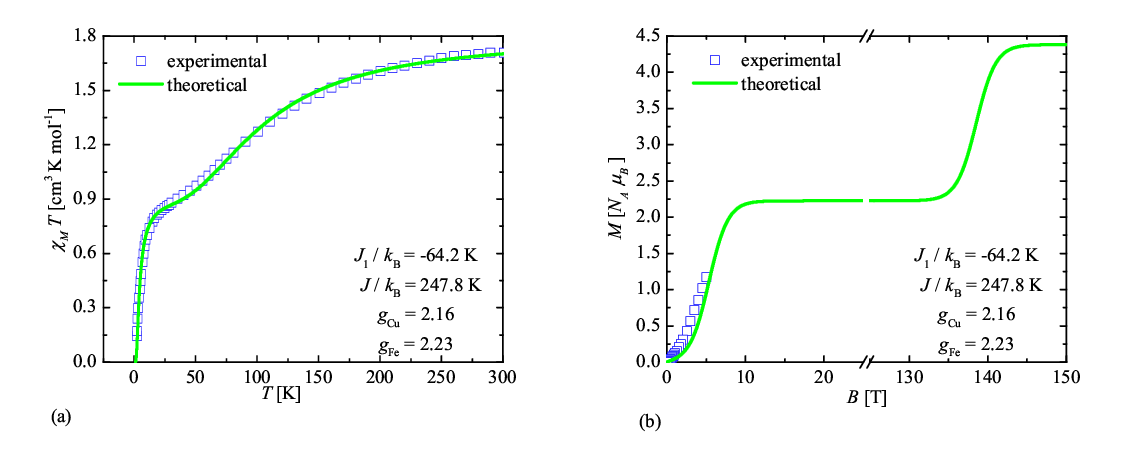}
\vspace{-0.7cm}
\caption{Temperature dependence of the molar magnetic susceptibility $\chi_{M}$ times temperature $T$ data at the magnetic field $B = 0.2$~T (a) and the isothermal magnetization curve of the polymeric coordination compound $[(\rm Tp)_{2}\rm Fe_{2}(\rm CN)_{6}$(CH$_3$COO)$(\rm bdmap)\rm Cu_{2}(\rm H_{2}\rm O)]$ recorded at temperature $T=1.8~\rm{K}$ (b). Blue squares represent experimental data 
reported in Ref. \cite{kan10}, while green curves are the respective theoretical fits obtained with the help of the spin-1/2 Ising-Heisenberg branched chain model.}
\label{experiment}
\end{figure*}

In this section, we will take advantage of the spin-1/2 Ising-Heisenberg branched-chain model described by the Hamiltonian \eqref{IsingHeiHam} to theoretically interpret available magnetic data experimentally measured for the heterobimetallic coordination polymer  $[(\rm Tp)_{2}\rm Fe_{2}(\rm CN)_{6}$(CH$_3$COO)$(\rm bdmap)\rm Cu_{2}(\rm H_{2}\rm O)]$ to be further abbreviated as Fe$_2$Cu$_2$, which serves as its experimental realization \cite{kan10}. The experimental data for the temperature dependence of the magnetic susceptibility times temperature product $\chi_{M} T$ recorded at a magnetic field of $B = 0.2$~T \cite{kan10} are confronted in Fig. \ref{experiment}(a) with the best theoretical fit obtained using the spin-1/2 Ising-Heisenberg branched-chain model. The least-squares fitting procedure yielded a rather strong antiferromagnetic coupling constant $J/k_{\rm B}=247.8$~K between nearest-neighbor Cu$^{2+}$-Cu$^{2+}$ magnetic ions, a moderately strong ferromagnetic coupling constant $J_{1}/k_{\rm B}=-64.2$~K between nearest-neighbor Cu$^{2+}$-Fe$^{3+}$ magnetic ions, and Land\'e $g$-factors $g_{\rm Cu}=2.16$ and $g_{\rm Fe}=2.23$ for Cu$^{2+}$ and Fe$^{3+}$ magnetic ions, respectively. A moderate value of the interaction ratio $J_1/J \approx -0.26$ contributes to a relatively rapid decline in the susceptibility times temperature product as temperature decreases in a high-temperature regime $T \gtrsim 50$~K. This is successively followed by a less steep decline observed in a range of moderate temperatures 20~K $\lesssim T \lesssim$ 50~K before it finally rapidly falls to zero with further temperature reduction. Note furthermore that the steep reduction of the susceptibility times temperature product observed in the low-temperature range is consistent with the quantum antiferromagnetic (QAF) ground state $|I'\rangle$.

Utilizing the determined set of fitting interaction parameters, we have also conducted theoretical modeling of the low-temperature magnetization curve of the polymeric compound Fe$_2$Cu$_2$, as reported in Ref. \cite{kan10}, for sufficiently low temperatures $T=1.8~\rm{K}$ up to a magnetic field of $B = 5~\rm{T}$. The magnetization data recorded for the coordination polymer Fe$_2$Cu$_2$ are confronted 
in Fig. \ref{experiment}(b) with the corresponding theoretical fit based on the spin-1/2 Ising-Heisenberg branched chain model, calculated up to an ultrahigh magnetic field of $150~\rm{T}$ surpassing the saturation field. The spin-1/2 Ising-Heisenberg branched chain model provides a relatively satisfactory fit to the experimental magnetization data. Moreover, an extrapolation to higher magnetic fields predicts the presence of a sizable intermediate one-half magnetization plateau in the magnetic field range between $B\approx 10~\rm{T}$ and $140~\rm{T}$. This intermediate magnetization plateau reflects the emergence of the quantum ferrimagnetic phase $|II'\rangle$ accompanied by a stronger bipartite entanglement of the Heisenberg spin pairs. It is worthwhile to mention that a small deviation between the theoretical and experimental data can be attributed to the Ising approximation of a highly (but not infinitely) anisotropic exchange coupling between Cu$^{2+}$ and Fe$^{3+}$ magnetic ions. 

\begin{figure}[t]
\centering
\vspace{-0.5cm}
\includegraphics[width=0.99\textwidth]{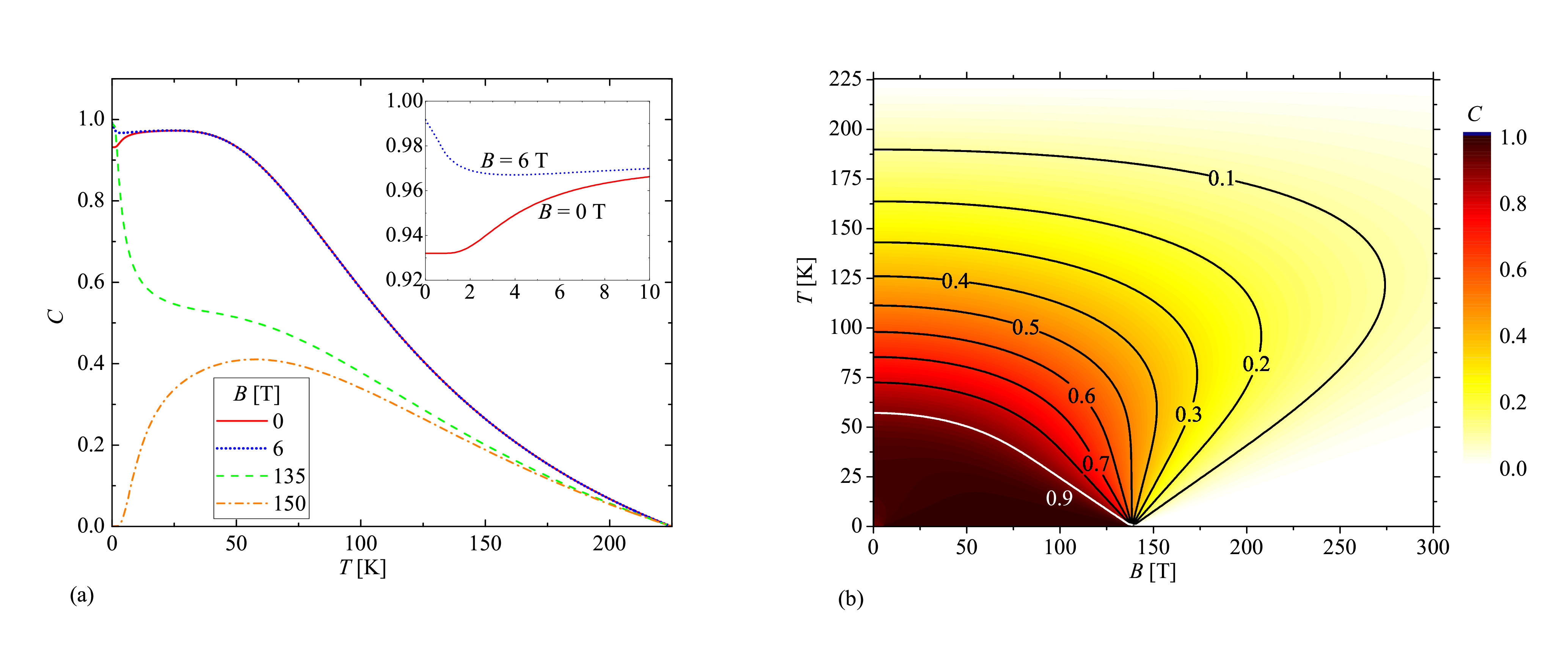}
\vspace{-0.8cm}
\caption{(a) Temperatures dependencies of the concurrence \eqref{Concurrence} calculated for the Cu$^{2+}$-Cu$^{2+}$ dimeric units using the spin-1/2 Ising-Heisenberg branched chain for several values of the magnetic field $B$ (the inset shows a detail from the low-field region); (b) A density plot of the concurrence \eqref{Concurrence} calculated for the Cu$^{2+}$-Cu$^{2+}$ dimeric units of the spin-1/2 Ising-Heisenberg branched chain in the magnetic field vs. temperature plane. Displayed lines correspond to isocontour lines of the concurrence. All calculations are presented for the set of fitting parameters $J/k_{\rm B}=247.8~K$, $J_{1}/k_{\rm B}=-64.2~K$, $g_{Cu}=2.16$ and $g_{Fe}=2.23$ obtained for the polymeric coordination compound $[(\rm Tp)_{2}\rm Fe_{2}(\rm CN)_{6}$(CH$_3$COO)$(\rm bdmap)\rm Cu_{2}(\rm H_{2}\rm O)]$.}
\label{predictionConcu}
\end{figure}

To shed light on the strength of bipartite entanglement within the Cu$^{2+}$-Cu$^{2+}$ dimeric units of the polymeric chain Fe$_2$Cu$_2$, we utilized the determined fitting parameters of the corresponding spin-1/2 Ising-Heisenberg branched chain to calculate the concurrence \eqref{Concurrence}. Temperature dependencies of the concurrence \eqref{Concurrence} calculated for the Heisenberg spin pairs of the spin-1/2 Ising-Heisenberg branched chain are illustrated in Fig. \ref{predictionConcu}(a) for a few selected values of the magnetic field $B$. At zero magnetic field, the concurrence exhibits a striking temperature-induced rise almost to its maximum possible value, after which it gradually declines with increasing temperature until it completely vanishes at the threshold temperature $T \approx 224~\rm{K}$. A small magnetic field can be considered as another driving force that may reinforce the strength of bipartite entanglement. Indeed, it turns out that the concurrence approaches almost its maximum value in the zero-temperature asymptotic limit when a small magnetic field is applied. If the magnetic field exceeds the saturation field, a double reentrant behavior of the concurrence is observed.

Last but not least, a density plot of the concurrence \eqref{Concurrence} calculated for the Heisenberg spin pairs of the spin-1/2 Ising-Heisenberg branched chain is presented in Fig. \ref{predictionConcu}(b) in the magnetic field versus temperature plane. This density plot can be interpreted as a phase diagram, delineating the entangled parameter region from the disentangled one. Remarkably, the density plot in Fig. \ref{predictionConcu}(b) reveals that a relatively high value of the concurrence ($C=0.1$) can still be detected even up to relatively high temperatures of around $T\approx 188~\rm{K}$ irrespective of the accessible magnetic-field strength for experimental testing ($B<10~\rm{T}$). This suggests the persistence of significant bipartite entanglement in the system over a wide temperature range, which could be of interest for experimental investigations.

\section{Conclusion}
\label{conclusion}

In this paper, we have investigated the ground-state phase diagram, magnetization process, magnetocaloric effect, and bipartite entanglement of the spin-1/2 Ising-Heisenberg branched chain with two different coupling constants using the transfer-matrix method. Our analysis reveals that the spin-1/2 Ising-Heisenberg branched chain exhibits five distinct ground states: two quantum antiferromagnetic phases $\left|I\right\rangle$ and $\left|I'\right\rangle$, two quantum ferrimagnetic phases $\left|II\right\rangle$ and $\left|II'\right\rangle$, and the classical ferromagnetic phase $\left|III\right\rangle$. These ground states are separated by discontinuous magnetic-field-driven phase transitions, whereby the ground states $\left|I,I'\right\rangle$ and $\left|II,II'\right\rangle$ are responsible for zero and one-half magnetization plateaus, respectively. Furthermore, our exact calculation of the concurrence has confirmed the quantum character of the ground states $\left|I,I'\right\rangle$ and $\left|II,II'\right\rangle$. We find that the strength of bipartite entanglement within the quantum ferrimagnetic phases $\left|II,II'\right\rangle$ is stronger compared to that within the quantum antiferromagnetic phases $\left|I,I'\right\rangle$. This insight contributes to a deeper understanding of the entanglement properties of the system and its magnetic behavior.

It has been demonstrated that the spin-1/2 Ising-Heisenberg branched chain with two different coupling constants provides a satisfactory description of the magnetic behavior observed in the heterobimetallic polymeric chain Fe$_2$Cu$_2$. By fitting available experimental data for the temperature dependence of the susceptibility times temperature product and the low-temperature magnetization curve, we have determined a rather strong coupling constant $J/k_{\rm B}=247.8~K$ between the Cu$^{2+}$-Cu$^{2+}$ magnetic ions and a moderately strong coupling constant $J_{1}/k_{\rm B}=-64.2~K$ between the Cu$^{2+}$-Fe$^{3+}$ magnetic ions with subtle differences in the Land\'e $g$-factors $g_{Cu}=2.16$ and $g_{Fe}=2.23$. Our analysis reveals that the quantum antiferromagnetic phase is realized at sufficiently low magnetic fields ($B \lesssim 10$~T), while the quantum ferrimagnetic phase, characterized by a sizable intermediate one-half magnetization plateau, can be detected in the magnetic-field range between $B\approx 10~\rm{T}$ and $140~\rm{T}$. The substantial coupling constant between the Cu$^{2+}$-Cu$^{2+}$ dimeric units further implies a robust bipartite entanglement between the nearest-neighbor Cu$^{2+}$-Cu$^{2+}$ magnetic ions. This entanglement persists up to a relatively high threshold temperature ($T \approx 224$~K) and undergoes a transient magnetic-field-driven strengthening.
 
\section*{Conflict of Interest Statement}
The authors declare that the research was conducted in the absence of any commercial or financial relationships that could be construed as a potential conflict of interest.

\section*{Author Contributions}
DS: software, investigation, formal analysis, methodology, data curation, validation, visualization, and writing–original draft. KK: conceptualization, investigation, formal analysis, methodology, data curation, validation, supervision, writing–review and editing. JS: conceptualization, investigation, formal analysis, funding acquisition, methodology, data curation, validation, visualization, supervision, project administration, writing–original draft, writing–review and editing. DS and JS contributed equally to this work. DS and JS share first and last authorship.

\section*{Funding}
This work was financially supported by the grant of the Slovak Research and Development Agency provided under the contract No. APVV-22-0172 and by the grant of The Ministry of Education, Science, Research, and Sport of the Slovak Republic provided under the contract No. VEGA 1/0695/23. This project has received funding from the European Union’s Horizon 2020 research and innovation programme under the Marie Sk\l odowska-Curie grant agreement No. 945380.



\section*{Data Availability Statement}
The datasets presented in this article are not readily available. The data are available upon reasonable
request from the corresponding author. 

\bibliographystyle{Frontiers-Vancouver}

\end{document}